# Prover and Verifier Based Password Protection: PVBPP

*Priyanka Naik[1*], Sugata Sanyal[2]*

*Computer Science Department, Manipal Institute of Technology, Manipal, India*

*ppnaik1890@gmail.com*

*Corporate Technology Office, Tata Consultancy Services, Mumbai, India*

*sugata.sanyal@tcs.com*

*\*Corresponding Author*

## Abstract

In today's world password are mostly used for authentication. This makes them prone to various kinds of attacks like dictionary attacks. A dictionary attack is a method of breaking the password by systematically entering every word in a dictionary as a password. This attack leads to an overload on the server leading to denial of service attack. This paper presents a protocol to reduce the rate of dictionary attack by using a prover and a verifier system. This system makes it difficult for the attacker to prove it as a valid user by becoming computationally intensive. The rate of attempts is also reduced and thus restricting the Denial of Service attack.

## 1. Introduction

In the computer and network security, the authenticated key exchange (AKE) protocol is employed to legitimatize the server and the client [1]. The user-memorable passwords along with username are commonly used to authenticate a user, as it is much simpler to implement and client friendly. Passwords are kept secret and are known only to valid user. But easier a password is for the owner to remember translates to easier for an attacker to guess it [2]. It makes the password vulnerable to dictionary attack and may lead to Denial of Service attack (DoS) [3]. The proposed protocol uses a fast one way hash function and reduces the rate of attempt with each wrong attempt [4]. The proposed protocol is completely stateless.

The rest of the paper is organised as follows:

In section 2 we discuss the related approaches that are being used currently with their merits and demerits. Section 3 deals with the proposed protocol. Section 4 shows some modification to the protocol and Section 5 consists of the conclusion and the future work.

## 2. Related Approaches

Earlier methods include Electronic Key Exchange (EKE), which uses the concept of shared keys [5]. In this method client and server share a common password which is a combination of asymmetric (public-key) and symmetric (secret-key) cryptography [6]. This method is secure against both online and offline attack to a great extent. However due to the usage the security efficiency of the protocol is compromised. Improvement was done by modifying 2-party Diffie-Hellman key exchange to n-party case [7]. As this protocol suffered from off-line and undetectable online guessing attack, modification was done using three-party key exchange protocols with password authentication [8]. An easily implementable challenge response system for eliminating online dictionary attacks was proposed [9].

### 2.1 Use of Strong Passwords

Making the password longer than eight characters and combining letters, numbers and special characters make a password strong [10]. The password can be made strong by using characters like @, $, %, #. An alternative way to build a strong password is to use the initial letters from a sentence and then shorten the words like converting "to" into 2, "s" into 5 etc. But such a password is difficult to remember; thus increasing the number of attempts for a valid user.

### 2.2 Generating Knowledgebase based on Password Usage

Keeping a track about the system which the user uses to login and detecting an attempt to login from an unknown system can be considered as a possible attack. The track of system can be kept by storing the Media Access Control address (MAC address) and Internet protocol (IP address) of system through which a user normally logs in. These addresses uniquely identify a system in the network. Automated individual white-list (AIWL) to keep track of the features of login pages (e.g. IP addresses, document object model (DOM) paths of input widgets) in the individual white-list can be used [11]. By checking the legitimacy of these features, AIWL can efficiently defend users against hard attacks, especially pharming and even dynamic pharming. This method leads to an overhead of storage at the server side as storing both the addresses requires a lot of memory.

## 2.3 Single Sign-On

With the growth of the networked world, users have many accounts in many services e.g. bank accounts, e-mail or chat accounts. After sometime, it becomes really difficult to manage so many passwords. Single Sign-On is a mechanism where all these passwords are kept in a software safe. All you need is a key or password to the safe. Once you sign on in this safe, all other systems open for you [12]. This means, with a single sign-in it is possible to access all the facilities. Single sign-on helps reducing human typographic error, but is difficult to implement as all the services need to be coordinated. If an attacker gets access to the software safe password it can access the entire information of the victim [13].

## 3. The Proposed Protocol

The proposed protocol reduces the rate of attack by increasing the delay for the password prompt with each failed attempt in a session. Each session is uniquely identified by using session ID. The key generated in each session are passed using secure channel like an SSL connection. SSL provides protection against network attacks by validating public-key certificates for each connection [14]. The protocol uses secure hash algorithm (SHA) for the one way hash function [4]. As the server does not store any password as a plain-text; the storage is also secured. A challenge response system is used to verify a user [9]. A block diagram of the proposed protocol is shown in Figure 1.

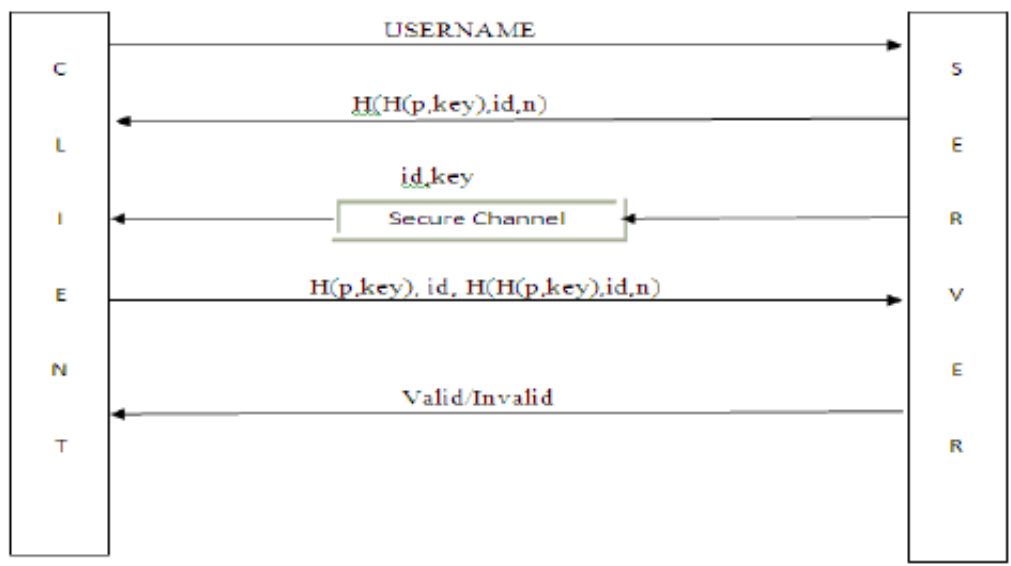

Figure 1. Messages exchanged between the client and server

p: password of the user

key: the random key generated

H(z): The one way hash function used

n: the number of attempts in this session

id: the unique session id generated by the server

**3.1 Explanation**

1. The client first sends the username to the server. This indicates the start of a session. The client system can be monitored for possible intrusion [15].
2. Upon receiving the username the server generates a unique session id and a random key. MAC is used to verify authenticity and integrity of a message [16]. The server generates the Message Authentication Code (MAC) i.e. H(H(p, key), id, n) and sends it to the client. Here n indicates the number of attempt i.e. for the first login attempt n is 1, which is incremented with each failed attempt. This value is used for increasing the delay with each attempt. The MAC is basically the hash value of the session id generated, n and the hash value of the stored password along with key. This can be sent through an insecure channel as it is a one way hashed value. The client does not use the MAC. It is used by the server itself to compare the client password. Sending the MAC prevents extra storage on server side. MAC is used for verification by the server.
3. Server also sends the session id and key generated to the client through a secure channel. The session id and the key are generated randomly by the server. Dynamic session id based system is used and improved to prevent masquerading attacks [17]. The id and key are the same which were used by the server to generate the MAC.
4. Client on receiving the MAC, id and key; hashes the password using the key and sends it to the server. This hash value can only be calculated by a valid user as he/she knows the key which is not known to the attacker. The hash function used is also the same one way hash function as used by the server.
5. The server compares the received hash value of the password from client with hash value it had calculated for the MAC. If the hashed values are same server declares the client as a valid client and sends a valid notification. If the values are not the same it is implied that either the client typed a wrong password or, it is an attacker trying to break the password.
6. If the response is invalid the client or attacker will try to enter the password again. But for the next attempt, the value of n will be incremented for that session. For successive attempt after the wrong attempt in that session, the session id remains the same but the value of n will be incremented. For each increment in n, the delay in showing the password prompt is increased exponentially. To achieve this, server method for this client's session is put to sleep for a certain period of time. It is this time which increases with each attempt. That is, if the delay to show the prompt on $1^{st}$ wrong attempt is 2 seconds, then for the next wrong attempt the delay will be 4 seconds. The delay implies that the client will not be shown any prompt to enter the password for the time for which the server sleeps for that client session. The delay time follows the value $2^n$, where n is the number of attempt. The base 2 is chosen to be compatible with the rest of the functions using binary values.

Thus, the attempt of an attacker to crack the password is delayed exponentially with each attempt. This proves useful as a valid user can type incorrect password once or twice but an attacker using dictionary attack will attempt login several times. Further the key is known only to a valid user and not to an attacker. Thus the hash can be correctly calculated by a valid user only. Here the client is the prover and the server is verifier [18]. The secure channel used can be an SSL connection between the client and server. The key is generated for each session and hence replay attack can also be prevented. The comparison of the proposed "Prover and Verifier based Password Protection" with CompChall protocol proposed by Goyal et al. [9] is given in Table 1.

## 4. Modification

Taking the estimated value of n as 10 million, the delay will be $2^{10000000}$ which will delay the prompt for a very long time [19]. Instead of taking the number of attempts for calculation of delay, timestamps can be used. Time at which the request received by the server and time at which the hashed value of password received from the client by the server can be recorded, and the time interval can be used to calculated the delay. The slower the client is to calculate the hash the larger will be its delay for the next attempt, if the login attempt fails. The timestamp reduces the calculation of delay value. Timestamp also helps to keep track of the time of login attempt which might help to trace down a password guessing attempt by an attacker. Keeping track of time at each intermediate node can be used for adding more security to the system [20]. To calculate the timestamp, the difference in inter arrival time can be used [21]. Time window can also be used for timestamp calculation [22]. The one way hash function used gives a throughput of 2 Gbps, which is very efficient [23]. The time required to calculate the hash value for a GHz system is 3 seconds [24]. This helps adding more delay for re-login after a failed attempt.

Table 1 Comparison of CompChall [9] and present protocol

| Parameter | CompChall protocol | Prover-Verifier protocol |
|---|---|---|
| Hash Decryption | Present on client side. | Not present. |
| Hash Comparison | Present on server side. | Present on server side. |
| Hash Generation | At client side using random "r" | At client side using the key. |

## 5. Conclusion and Future Work

In the present paper, Prover and Verifier Based Password Protection protocol is used to eliminate online dictionary attacks. The proposed protocol reduces the rate of attack by delaying the password prompt exponentially with each failed attempt in a session. In the protocol, on receiving the request from client with username, the server generates a unique session id and a random key and generates the Message Authentication Code and sends it to client through an insecure channel, as it is a one way hashed value. Server also sends the session id and key to the client through a secure channel. Client hashes the password using the key and sends it to the server. The server verifies the received hash value of the password from client by matching with the hash value calculated for the MAC and checks the authenticity of the client. To make multiple attempt time consuming, a prompt to enter the password is delayed for each failed attempt exponentially, by putting the server on sleep mode for a calculated delay. The protocol can be modified by inclusion of timestamp for the server response. Timestamps in addition to delaying the prompt response also help in keeping the track of the time of login attempt, which might help to trace down a password guessing attempt by an attacker.